\documentclass{article}
\usepackage{spconf,amsmath,graphicx}
\usepackage{xcolor}
\usepackage{multirow}
\usepackage{comment}
\usepackage{url}
\usepackage{hyperref}
\usepackage{enumitem}
\usepackage{enumitem}
\setlist{nosep, leftmargin=14pt}
\usepackage[tableposition=above]{caption}

\usepackage{mwe} %

\title{Data-Centric Label Smoothing for Explainable \\Glaucoma Screening from Eye Fundus Images}

\name{Adrian Galdran$^{1,2}$ \qquad Miguel A. González Ballester$^{2,3}$}
 \address{$^{1}$ Computer Vision Center, Universitat Autònoma de Barcelona, Spain \\$^{2}$ BCN Medtech, Dept. of Information and Communication Technologies,\\ Universitat Pompeu Fabra, Barcelona, Spain\\$^{3}$ ICREA, Barcelona, Spain}

\begin{document}
\maketitle
\begin{abstract}
As current computing capabilities increase, modern machine learning and computer vision system tend to increase in complexity, mostly by means of larger models and advanced optimization strategies. 
Although  often neglected, in many problems there is also much to be gained by considering potential improvements in understanding and  better leveraging already-available training data, including annotations. This so-called data-centric approach can lead to substantial performance increases, sometimes beyond what can be achieved by larger models. 
In this paper we adopt such an approach for the task of justifiable glaucoma screening from retinal images. 
In particular, we focus on how to combine information from multiple annotators of different skills into a tailored label smoothing scheme that allows us to better employ a large collection of fundus images, instead of discarding samples suffering from inter-rater variability. 
Internal validation results indicate that our bespoke label smoothing approach surpasses the performance of a standard resnet50 model and also the same model trained with conventional label smoothing techniques, in particular for the multi-label scenario of predicting clinical reasons of glaucoma likelihood in a highly imbalanced screening context. 
Our code is made available at \url{github.com/agaldran/justraigs} .
\end{abstract}

\section{Introduction and Related Work}
Since the introduction of the transformer architecture \cite{vaswani_et_al_attention_2017}, many recent advances in the area of machine learning have resulted from the optimization of scaling up models and improving learning algorithms, but the field has also realized that it is equally important to handle training data effectively.
Enhancing the quality and quantity of training data by engineering it before moving it into machine learning systems is an area of research known as data-centric artificial intelligence.
As opposed to standard model-centric approaches, where we focus on designing better learning mechanism and models to improve performance, data-centrism investigates possible deficiencies in data, \textit{e.g.} missing values \cite{dalca_et_al_medical_2019}, cleaning wrong labels \cite{northcutt_confident_2021}, or finding and removing out-of-distribution samples prior to training \cite{nayak_comprehensive_2023}. 
In computer vision, applications range from semantic segmentation \cite{galdran_et_al_no-reference_2018} or object detection \cite{tkachenko_objectlab_2023}.
A recent review on the topic can be found in \cite{zha_et_al_data-centric_2023}.

In the medical image analysis field, the maximum exponent of data-centric strategies is arguably the widely popular nnU-net segmentation framework \cite{isensee_nnu-net_2021}.
In this case, given a 3d medical image dataset, the system automatically generates a footprint indicating not only the type of CNN architecture to be used, but also a highly performing set of hyperparameters, e.g. volumetric patch size, batch size, pre-processing operations, and so on. 
Designed to compete in the Medical Decathlon challenge \cite{antonelli_et_al_medical_2022}, the nnU-Net has since become the default baseline over which to build improvements on 3d medical segmentation tasks \cite{sundar_et_al_fully_2022,wasserthal_et_al_totalsegmentator_2023,mcconnell_exploring_2023,isensee_et_al_extending_2023}.

In this article, we adopt a data-centric approach for the task of explainable glaucoma screening from retinal fundus images. 
Glaucoma is a sight-threatening disease that represents the second leading global cause of blindness, impacting over 91 million people worldwide \cite{thakur_predicting_2020,madadi_domain_2024}. 
Due to the ease of acquisition and wide availability of public retinal image fundus data collections, a large set of classification models for the detection of glaucoma from fundus images have been proposed in recent years\cite{hemelings_et_al_generalizable_2023}, including in the context of public competitions \cite{orlando_et_al_refuge_2020}. 
However, the black-box nature of deep neural networks has prevented these high-performing models from reaching common clinical practice\cite{wu_performances_2022}. 
Some authors have attempted to predict and regress common biomarkers such as vertical cup-to-disc ratio \cite{hemelings_et_al_deep_2021}, or more recently the Rim Thickness Curve \cite{wundram_et_al_leveraging_2024}.
Another route towards greater explainability of glaucoma diagnosis is by means of directly predicting a set of relevant clinical features that result in a clinician declaring disease presence. 
Even if this might be an optimal solution, it requires richly annotated data that has only very recently been available to the community\cite{lemij_characteristics_2023}.

In this work, we focus on the engineering of a tailored label smoothing that reflects inter-rater disagreement. 
Our system is designed to take part in the JustRaigs competition, where the provided dataset \cite{lemij_characteristics_2023} featured a large scale set of retinal images labeled by a pool of annotators of varying expertise. 
We quantitatively show that incorporating information on the amount of expertise into soft labels can enhance the predictive ability of a standard ResNet50 model for the tasks of glaucoma screening and justification.

\begin{figure*}[!t]
\centerline{\includegraphics[width=0.85\textwidth]{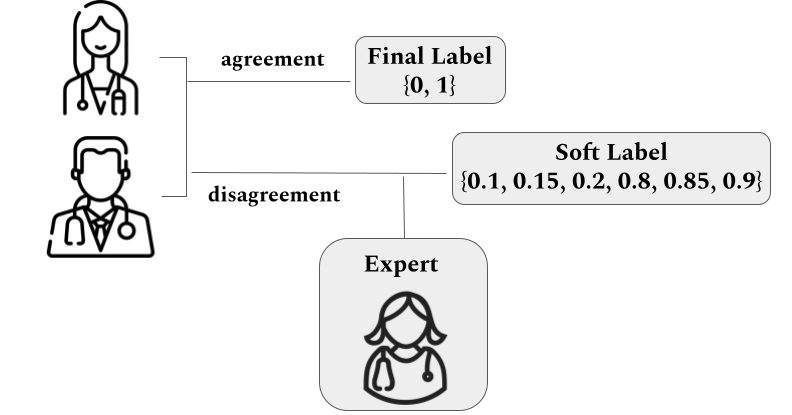}}
\caption{Data-Centric Label Smoothing: depending on the sort of disagreement present on annotations, and the skill of the involved annotators, the degree of smoothing appliced to labels will vary.}
\label{fig_overview}
\end{figure*}

\section{Methodology}
In this section, we first give an overview of our baseline model, composed of a standard computer vision architecture trained with conventional techniques. 
We then describe the main features of the dataset we used in this paper, focusing on the nature and quality of its annotations. The last subsection explains how we incorporated this knowledge into a specific label smoothing scheme suitable for our learning task.

\subsection{Baseline Model}
In this paper, we considered a ResNet50 architecture, which has been shown as an extremely competitive computer vision model with a good compromise between complexity and accuracy \cite{wightman_resnet_2021}. 

Since we intend to take a data-centric approach, we do not spend much time optimizing hyper-parameters. This is, we optimize the network with an Adam algorithm and a default learning rate of $l=1e-4$, batch-size of 8, minimizing a regular Cross-Entropy loss for as long as we do not detect overfitting on a separate validation set comprising 20\% of the data (we run a 5-fold training ensemble). 
We crop the images to their field of view, and apply common data augmentation strategies, \textit{e.g.} random flipping and rotations, or small image intensity perturbations.

\subsection{Data and Annotations}
The dataset provided by the competition organizers contains 113,893 retinal images labeled for glaucoma analysis. 
Specifically, there is a main annotation reporting the status of glaucoma in the patient, but also a subset of the images has a rich collection of supplementary annotations, as explained next.

In order to understand our approach, it is important to explain the details of the annotation process. 
Initially, a pool of graders, both ophthalmologists and optometrists, graded the images. 
For each image, a randomly selected pair of raters reported the image to be with referable glaucoma (RG), no referable glaucoma (NRG), or ungradable (U). 
Whenever there was disagreement between grader 1 (G1) and grader 2 (G2), a glaucoma expert (G3) resolved the grading. 
On the other hand, performance of graders was monitored and part of them abandoned the study if their accuracy was not deemed sufficient. 
Their images were reannotated, unless G3 had already gave a diagnosis, in which case the low-quality diagnosis was simply removed. 
As a consequence, a small subset of the labels for G1 or G2 would become unavailable (NaN).
Eventually, in order to reach a binary label, an annotation considered as final whenever there is agreement between G1 and G2, or in case of disagreement, the final label is the one provided by by the specialist G3.

For the explainability task, images identified as showing signs of RG were given annotations for 10 clinically relevant glaucomatous features $f_1, ..., f_{10}$ \cite{lemij_characteristics_2023}. 
In this case, if both G1 and G2 initially reported RG, then the dataset contains two sets $f^1_1, ..., f^1_{10}$ and $f^2_1, ..., f^2_{10}$ that may not be equal, but there was no adjudication in cases of disagreement at the glaucoma feature level. 
In addition, if the image was deemed glaucomatous by one of both graders, and G3 agreed, then we also have two sets of feature values that may not coincide. 
For purposes of evaluation, the competition organizers discarded any feature values that showed disagreement.

A straightforward approach would consist of training a model using only final referrable glaucoma annotations and discarding all information regarding disagreement. 
In contrast, we attempt to incorporate this into the labels used for training, as explained in the next subsection.

\subsection{Multi-Rater Data-Centric Label Smoothing}
Conventional Label Smoothing strategies for binary classification attempt to regularize the training process by substituting ``hard labels'' $y\in\{0,1\}$ for hard-coded soft values $\tilde{y}\in\{0.1,0.9\}$, thereby preventing a network trained with cross-entropy loss from becoming overly confident \cite{galdran_multi-head_2023,murugesan_calibrating_2023}.

We propose label smoothing to encode inter-rater disagreement by following a set of rules described below:
\begin{itemize}
\item Whenever the final label $y=0$ or $y=1$ but a rater considered the image as ungradable ($U$), we use soft labels $\tilde{y}=0.1$ or $\tilde{y}=0.9$ instead.
\item Whenever there was disagreement between $G_1$ and $G_2$, instead of directly adopting the decision of $G_3$, we soften it so that if $G3=G_i=0, G_j=1$, we use $\tilde{y}=0.15$; conversely, if $G3=G_i=1, G_j=0$, we use $\tilde{y}=0.85$.
\item If an annotation is missing, \textit{i.e.} $G_1$=NaN, and $G_2!=G_3$, instead of considering $G_3$ as the final decision, if $G_3=0$ then we use $\tilde{y}=0.2$ and if $G_3=1$, we consider $\tilde{y}=0.8$.
\end{itemize}

\noindent For the feature-level labels, we have sets of ten binary annotations, that we pre-process as follows:
\begin{itemize}
\item If $G_i=1\neq G_j$, with $G_3=1$, we have two sets of feature values. 
When there is disagreement at the feature level, we assign $f\in\{0.1,0.9\}$ favoring the opinion of G3.
\item If $G_i=1\neq G_j$, with $G_3=0$, we have one set of feature values. 
For each feature that has a value of $f=1$, we consider a label of $f=0.25$
\item This still leaves cases where $G_i=G_j=1$, but not at the feature level. In case of disagreement at the feature level in this context, we use $f=0.5$.
\end{itemize}

The resulting set of different smooth labels we use is illustrated in Fig. \ref{fig_overview}.

\begin{table}[!h]
\renewcommand{\arraystretch}{1.3}
\begin{center}
\begin{tabular}{c c c c c c} 
  & Fold 1 & Fold 2 & Fold 3 & Fold 4 & Fold 5\\ [0.5ex] 
 \hline
 Final & 93.12 & 92.05 & 92.97 & \textbf{94.33} & 92.51 \\ 
 \hline
 LS    & 92.97 & 91.59 & 92.35 & 93.57 & 90.67\\
 \hline
 DC-LS & \textbf{93.27} & \textbf{93.43} & \textbf{93.27} & 93.87 & \textbf{92.6}6\\
 \hline
\end{tabular}
\end{center}
\caption{Five fold sens@95spec for Glaucoma screening using different label configurations. Best results \textbf{boldfaced}.}
\label{t1}
\end{table}

\section{Experimental Results}
We trained the same model five times by separating the training and validation sets in 80/20 proportions\footnote{Code is available at \url{github.com/agaldran/justraigs}}. 
In Table \ref{t1} we show, for the referrable glaucoma problem (binary classification) the results of these experiments when using only data with final decisions (Final), when using all data but a uniform label smoothing (LS), and when using the data-centric label smoothing (DC-LS) scheme described in the previous section. The metric of choice is sensitivity at 95\% specificity, as indicated in the competition guidelines.

We can see how the data-centric label smoothing results in improvements in terms of sensitivity in all folds but one. 
In addition, even when standard label smoothing leads to degraded performance, our approach still improves the model's results.
This insight is confirmed in Table \ref{t2}, where we report results for the explainable glaucoma feature prediction task, terms of Hamming error between the true feature vector and the predicted one. We again compare against training on data with full agreement or expert opinion, plus when using label smoothing. 
We again see that data-centric label smoothing results on lower Hamming losses, indicating that the extra training data and the adapted soft labels can improve performance of a standard classifier also on this multi-label problem. 

\begin{table}[!t]
\renewcommand{\arraystretch}{1.3}
\begin{center}
\begin{tabular}{c c c c c c} 
  & Fold 1 & Fold 2 & Fold 3 & Fold 4 & Fold 5\\ [0.5ex] 
 \hline
 Final & 0.2251 & 0.2286 & 0.2127 & 0.2129 & 0.2352 \\ 
 \hline
 LS    & 0.1823 & 0.1764 & 0.1748 & 0.1743 & 0.1799\\
 \hline
 DC-LS & \textbf{0.1468} & \textbf{0.1440} &\textbf{ 0.1455} & \textbf{0.1488} & \textbf{0.1476}\\
 \hline
 \end{tabular}
 \end{center}
 \caption{Five fold Hamming Loss for Glaucoma features prediction with different label schemes. Best results \textbf{boldfaced}.}
\label{t2}
\end{table}

\section{Conclusion}
In this paper we describe Data-Centric Label Smoothing, an adaptation of conventional label smoothing that takes into account multi-rater disagreement s and different expert skills in order to define a set of soft labels, both for binary and a multi-label classification tasks. 
Our experiments show that introducing otherwise discared data with soft labels into the training of a standard Resnet50 model leads to substantial performance increases.

\section{Compliance with ethical standards}
This study was conducted retrospectively using data made available by the JustRaigs challenge in an open access Zenodo repository 
\href{https://zenodo.org/records/10035093}{(link)}.
Ethical approval was not required as confirmed by the license attached with the open access data.

\section{Acknowledgments}
\label{sec:acknowledgments}
A. Galdran was funded by MSC Fellowship No. 892297.

\bibliographystyle{IEEEbib}
\bibliography{datacentric_glaucoma}

\end{document}